# EXPERIMENTAL CHALLENGES INVOLVED IN SEARCHES FOR AXION-LIKE PARTICLES AND NONLINEAR QUANTUM ELECTRODYNAMIC EFFECTS BY SENSITIVE OPTICAL TECHNIQUES

by

Christopher C. Davis[1*], Joseph Harris[2], Robert W. Gammon[3], Igor I. Smolyaninov[1] and Kyuman Cho[4]

[1]*Department of Electrical and Computer Engineering University of Maryland, College Park, MD 20742*
[2]*Department of Physics, University of Maryland, College Park, MD 20742*
[3]*Institute for Physical Science and Technology, University of Maryland, College Park, MD 20742*
[4]*Department of Physics, Sogang University, Seoul, Korea*

We discuss the experimental techniques used to date for measuring the changes in polarization state of a laser produced by a strong transverse magnetic field acting in a vacuum. We point out the likely artifacts that can arise in such experiments, with particular reference to the recent PVLAS observations and the previous findings of the BFRT collaboration. Our observations are based on studies with a photon-noise limited coherent homodyne interferometer with a polarization sensitivity of $2 \times 10^{-8}$ rad Hz$^{1/2}$ mW$^{-1/2}$.

P.A.C.S. numbers: 42.15.Eq, 14.80.Mz, 12.20-m, 13.40.-f

*Introduction*

The most important magneto-optical interactions that can occur in material media are the Faraday effect, magnetic dichroism, and magnetic birefringence (the Cotton-Mouton effect). Quantum electrodynamics predicts that because of photon-photon interactions even the vacuum becomes birefringent in the presence of a strong magnetic field [1-5]. Further, the interaction with an axion-like particle and two photons via the Primakoff effect will also lend optical properties to the vacuum in the presence of a strong magnetic field [6-10]. The occurrence of an apparent magnetic dichroism of the vacuum would imply the preferential disappearance of left- or right circularly polarized photons from a light beam. To conserve mass and energy this would imply either the production of particles, or photon-splitting.

The QED effect and the axion effect are treated in terms of an effective Lagrangian [1-7], in units where $\hbar = c = 1$ and $\alpha = e^2/4\pi \approx 1/137$.

$$L = -\frac{1}{4}F_{\mu\nu}F^{\mu\nu} + \frac{\alpha^2}{90m_e^4}\left[\left(F_{\mu\nu}F^{\mu\nu}\right)^2 + \frac{7}{4}\left(F_{\mu\nu}\tilde{F}^{\mu\nu}\right)^2\right] + \frac{1}{2}\left(\partial_\mu a \partial^\mu a - m_a^2 a^2\right) + \frac{1}{4M}F_{\mu\nu}\tilde{F}^{\mu\nu}a \qquad (1)$$

Where the first half of the expression is the Euler-Heisenberg effective Lagrangian, which is appropriate to the QED effect, and the second half is the effective Lagrangian, which is appropriate to the Primakoff effect and accounts for the axion. Here, $a$ is the axion field, $m_a$ is the axion mass, and $M$ is the inverse axion coupling constant. Raffelt and Stodolsky [7] synthesize the results of Adler [4] and solve for the equations of motion. Analysis of the classical wave solutions of the equations of motion produces a picture of mixing between photon and axion modes in a polarized laser experiment with a static transverse magnetic field and an optical cavity to increase path length. In such an experiment, CP arguments predict that the axion will only couple to the parallel components of the beam. Thus, two main effects are predicted. The first effect is a phase difference $\Delta\phi = \phi_{//} - \phi_\perp$ between the parallel and perpendicular components of polarized light interacting with the magnetic field. This arises from both QED and the preferential mixing of axion and photon modes. In the mixing part of this picture, a photon mode oscillates into an axion mode before turning back into a photon and gets out of phase. In both cases, this phase difference causes an apparent birefringence.

The second main effect, is an apparent linear dichroism which manifests itself as a rotation, $\psi$, of the polarization and attenuation. This is caused by the fact that mirrors do not reflect axions and, hence, any axion modes that do not oscillate back to photons before hitting the mirror will appear as lost parallel photon modes. For small axion masses, the theory predicts:

$$\Delta\phi_{QED} = \frac{2\alpha^2 B_{ext}^2}{15m_e^4}\omega L, \quad \Delta\phi_a = N\frac{B_{ext}^2 m_a^2}{48M^2\omega}l^3 \text{ and } \psi_a = N\frac{B_{ext}^2}{8M^2}l^2, \qquad (2)$$

where $l$ is the length of the cavity, $N$ is the number of passes and $L = Nl$ is the total path length of the beam through the interaction region. The subscripts *QED* and *a*, refer to the origin of the effects. In terms of index of refraction, $\Delta\phi = kL(n_{//} - n_\perp)$. Choosing the limit of small axion masses is justified by several experimental results and astrophysical observations [6-11] which bound the axion mass to $10^{-3}eV > m_a > 10^{-6}eV$ and $M > 10^{10} GeV$. This result also takes into account Adler's analysis of the E-H Lagrangian which predicts the following vacuum birefringence:

$$n_\perp = 1 + 2\xi \sin^2 \theta, \; n_{//} = 1 + \frac{7}{2}\xi \sin^2 \theta \; \text{with} \; \xi = \frac{\alpha^2 B_{ext}^2}{45\pi m_e^4}. \tag{3}$$

Here, $\theta$ is the angle between $\mathbf{B}_{ext}$ and $\mathbf{k}$ .[4,7]

The expected birefringence, as a function of $B_{ext}$ in Tesla, due to the QED effect, is $\Delta n = n_{//} - n_\perp = 4 \times 10^{-23} B_{ext}^2$. The phase shift between the two orthogonal components of a light beam is $\Delta \phi = 2\pi L \Delta n / \lambda$. For input light linearly polarized at $45^o$ to the field direction, this translates into an induced ellipticity of the light $\varepsilon = \pi L \Delta n / \lambda$, for a path length $L$. For a 1m path and a 1T field the induced ellipticity is expected to be $1.2 \times 10^{-16}$. No experiment to date has achieved this sensitivity.

*The BRFT and PVLAS Experiments*

Two important experiments have attempted to detect the phenomena that would result from the Primakoff effect. In the BRFT experiment [12] an upper limit of $3.5 \times 10^{-10}$ rad was determined for the possible rotation angle for a 2.2km path in a 3.25T field, equivalent to $1.5 \times 10^{-14}$ rad m$^{-1}$T$^{-2}$, and an ellipticity of $1.6 \times 10^{-9}$ was measured on a 299m path in a 3.25T field. The PVLAS experiment [13] claims a rotation of $1.7 \times 10^{-7}$ rad for a 44km long path in a 5T field, equivalent to $1.55 \times 10^{-14}$ rad m$^{-1}$T$^{-2}$. The BRFT and PVLAS experiments differ in several important specific ways, although from the standpoint of applying a modulated magnetic field they are similar. BRFT uses a transverse magnetic field modulated at a frequency of 32mHz about a background level of 3.25T. PVLAS uses a transverse magnetic field that rotates around the light propagation axis at 1.89rad/s. This field is equivalent to the simultaneous application of two orthogonal transverse field components oscillating at 0.3Hz, but in quadrature. Neither the BRFT nor PVLAS experiments operated at the photon noise limit. The BRFT experiment used a 200mW argon ion laser and achieved a sensitivity of $4.7 \times 10^{-7}$ rad Hz$^{\frac{1}{2}}$ m W$^{-1/2}$. The PVLAS experiment used a 100mW 1.06µm Nd:YAG laser and achieved a sensitivity of $10^{-6}$ rad Hz$^{-1/2}$ mW$^{-1/2}$. The photon noise limit at 1.06µm for a detector with a responsivity of 0.4 A/W (a typical value for a Si photodiode at this wavelength) is $2 \times 10^{-8}$ rad Hz$^{1/2}$ mW$^{1/2}$.

*Discussion*

We have for several years operated a balanced coherent homodyne polarization interferometer for the study of the Faraday and Cotton-Mouton effects in condensed matter [14], and have achieved a photon noise limited sensitivity of $2 \times 10^{-8}$ rad Hz$^{-1/2}$ mW$^{-1/2}$ at 632.8nm or 1.06µm. Because we have only a 1kGauss modulated transverse field magnet with 0.1m pole pieces we could not compete with the BRFT and PVLAS experiments in overall sensitivity since we were a factor of $2.3 \times 10^4$ mT$^2$ below BRFT and a factor $10^9$ mT$^2$ below PVLAS in terms of path length and field strength. However, our experience with a very sensitive system for measuring elipticity has taught us much about the potential pitfalls of these experiments from an experimental optics standpoint. It is clear to us that the PVLAS experiment suffers from artifacts, as has already been pointed out by Melissinos [15], that the BRFT experiment suffers from artifacts has been acknowledged by its authors, although they do not specify all the sources of these spurious signals. A primary source of spurious signals in sensitive experiments of this kind is motion of optical components caused by a time-varying or a rotating magnetic field. The BRFT experiment acknowledges this and used a feedback system to attempt to minimize its effects. The PVLAS data show clear sideband peaks corresponding to the rotation frequency of their magnet, which should not be present for an effect proportional to $B^2$. Indeed these peaks are approximately 18 times larger than the "real" signal at twice the magnet rotation frequency. They do not explain the origin of the fundamental signal but interpret the second harmonic signal as resulting from an interaction involving a light, neutral, spin-zero particle.

In both the BRFT and PVLAS experiments optical components are either close to the magnet or mechanically coupled to the magnet and its cryostat. A primary component of the experiment that is strongly affected by the magnetic field is the evacuated tube passing though the magnet. This tube extends to the cavity end mirrors. All components in the experiment that experience any modulated field or field gradients will experience time-varying diamagnetic or paramagnetic forces. For example, any stainless steel or aluminum optical mounts will experience paramagnetic forces. There are torques acting on induced magnetic dipoles, especially in any components exposed to the field that are not absolutely symmetrically placed with respect to the field direction. A quartz sample tube in the magnet will experience the strongest forces in the regions where it leaves the magnet and experiences the largest field gradients, and will be pulled into the magnet bore. In general time-varying forces all result from any changes in magnetic stored energy that occur as the field is modulated. This generalized force on an object is $\mathbf{F} = -\nabla \int \frac{1}{2} \mathbf{B} \cdot \mathbf{H} dV$.

In our sensitive magneto-optical experiments we have verified that significant artifacts can result from any modulated feedback of light into the laser [16]. It has been shown that if a part of its own field is fed back into a laser by an optical component vibrating with small amplitude, then in the weak feedback regime, phase and amplitude of the output beam from the laser are synchronously modulated [17]. This effect is so efficient that when the source laser is influenced by the feedback the modulated light can cause interference in a sensitive measurement even for a balanced homodyne interferometer measuring an extremely small signal. We have performed a rigorous study of the feedback effect for the case of a balanced homodyne polarization interferometer. As a result, we have been able to detect phase and/or amplitude modulation produced in a balanced homodyne polarization interferometer when light from a mirror oscillating with an amplitude of only 9nm is fed back into the laser with 120dB of attenuation. This effect is still present even if the laser is an extremely low phase noise Nd:YAG ring laser [17]. The BRFT experiment is less sensitive to this feedback effect because it uses a multipass, zig-zag Herriott type cavity [18,19] rather than a spherical Fabry-Perot cavity. It is possible for light scattered by any of the optical components in these experiments to cause feedback, even if no specific optical component is used in the normal direction, and this includes scattered light that reflects off the inside walls of the evacuated tube inside the magnet. The BRFT experiment uses a single optical isolator, which probably does not provide sufficient isolation to prevent feedback modulation effects. It appears that, according to the experimental arrangement shown in ref [11], the PVLAS experiment does not use an optical isolator after its laser. In principle, the Fabry-Perot resonator might not reflect significant incident light if the source laser is perfectly frequency locked to the resonator. In practice, however, even for a very high-Q resonator, it is impossible to avoid the feedback due to imperfectness of mirrors and locking electronics. Therefore, in the PVLAS experiment, the feedback modulation effects may cause major interference in measurements.

In principle, any correlated intensity noise can be rejected in a balanced homodyne interferometer. However, because of the imperfect performance of real optical and/or electronic components, overall common mode rejection ratio of the interferometer used in our study was approximately 40dB. Synchronous feedback can cause interference in a sensitive experiment even when the signal level is very low. In the case of the PVLAS scheme, by including the feedback effect synchronized at twice the rotating frequency of the magnet, the representation for the light intensity transmitted through the crossed polarizers of the ellipsometer given in Eq (2) of Ref. [11] can be rewritten as

$$I = (I_0 + I_{N,2\nu_m})\{\sigma^2 + [\alpha(t) + \eta(t) + \Gamma(t)]^2\}$$

where $I_0$, $\sigma^2$, $\alpha$, $\eta$, and $\Gamma$ have the same meaning as in Ref [11] and $I_{N,2\nu_m}$ is the intensity modulation caused by the feedback. The frequency of this synchronized modulation is given by the vibration frequency of a feedback element, twice the frequency of the rotating magnet. Small misalignment between the polarization components must be included in the quasi-static, uncompensated rotation and ellipticity, $\Gamma$, which is much larger than the rotation caused by the Primakoff effect. Thus the term $2I_{N,2\nu_m}\eta(t)\Gamma(t)$ in the above equation has not only the same Fourier frequency as $2I_0\alpha\eta$ but also has the same phase relationship when the quarter-wave plate is rotated by 90$^o$. The synchronous interference, thereby cannot be distinguished from the magneto-optical effect being sought.

An important, but subtle distinction between the BRFT and PVLAS experiments is that the BRFT uses a mode-matched mirror cavity while the PVLAS apparently does not. Consequently, in the PVLAS experiment as the light beam oscillates between the two cavity mirrors its spot size and radius of curvature both oscillate and the radius of curvature does not match the mirror curvatures. This mismatch in radius leads to local non-normal incidence on the cavity mirrors (except on axis) and causes the local P-and S-polarization components of the beam to suffer different phase shifts, which vary radially on the mirror. A calculation for a typical very high reflectance multilayer mirror shows that this phase difference can be easily $10^{-11}$ rad per reflection for an incidence angle of 1.5mrad. The PVLAS cavity is subject to these effects, which would be modulated if the cavity mirrors move, although the BRFT cavity is not.

A potential confounder in a search for vacuum magneto optic effects is the Faraday effect resulting from residual axial field components and trace gas. There are residual axial field components in both the BRFT and PVLAS experiments, since the local wave-vector directions in a Gaussian beam are only nominally perpendicular to a transverse field at the beam waist, or on axis. We do not however, believe that these were the sources of sidebands at the magnet oscillation or rotation frequency ω$_m$. Nonetheless, an experiment in which there is no obvious modulation of the effect at frequency ω$_m$ is desirable, since an effect proportional to $B_{ext}^2$ only shows up at frequency $2\omega_m$. In an experiment in which the entire field is modulated at frequency $\omega_m$ a Faraday effect signal, or spurious signal, at $\omega_m$ is distinguished from the desired signal at frequency $2\omega_m$, which should be further checked by verifying that the desired signal is proportional to $B_{ext}^2$. A complication can arise if the magnet modulation is not a pure harmonic at frequency $\omega_m$. Any second harmonic of the magnetic field can produce a spurious signal at $2\omega_m$, but this can be identified since it will be linear in $B_{ext}$.

*Features of an Improved Experiment*

It is our belief that a balanced coherent homodyne interferometer is a better instrument to use than an extinction-based ellipsometer in a search for vacuum magneto-optical effects. Such a system is almost guaranteed to achieve the photon noise limit and provides excellent common mode rejection of laser noise. We also believe that any effect observed should be demonstrated to scale with $B_{ext}^2$ [14]. It will also be desirable to use the largest magnetic field possible, but not to modulate this. An experiment similar to PVLAS can then be performed by rotating the optical train at angular frequency $\omega_m$.

*Conclusions*

We believe that we have identified the likely causes of artifacts in the PVLAS experiment, and therefore suggest that the case for an interaction involving an axion-like particle has not been made. Furthermore, the PVLAS experiment contradicts the findings of the BRFT experiment, and a series of astrophysical observations that restrict the range of axion particle masses that are possible. An improved experimental arrangement is needed to pursue vacuum magnetic birefringence and polarization rotation effects. With an improved system, detection of the QED- predicted magnetic birefringence [4,5] should be possible, and a more sensitive examination of the existences of any axion-like interactions.

\* Corresponding author
Email address: davis@umd.edu